\documentclass{SciPost_better_arXiv}

\hypersetup{
    colorlinks,
    linkcolor={red!50!black},
    citecolor={blue!50!black},
    urlcolor={blue!80!black}
}

\usepackage[bitstream-charter]{mathdesign}
\usepackage{booktabs}
\usepackage[export]{adjustbox}
\usepackage{subcaption}

\DeclareSymbolFont{usualmathcal}{OMS}{cmsy}{m}{n}
\DeclareSymbolFontAlphabet{\mathcal}{usualmathcal}

\fancypagestyle{SPstyle}{
\fancyhf{}
\lhead{\colorbox{scipostblue}{\bf \color{white} ~SciPost Physics Community Reports }}
\rhead{{\bf \color{scipostdeepblue} ~Submission }}

\fancyfoot[C]{\textbf{\thepage}}
}

\RequirePackage{heppennames2}
\RequirePackage{lhchiggs}

\newcommand{\GOSAM}{{\tt{GoSam}}}
\newcommand{\WHIZARD}{{\tt{Whizard}}}
\newcommand{\GOWIZ}{{\tt{GoSam+Whizard}}}

\newcommand{\PROVBFHH}{\texttt{proVBFHH}\xspace}

\newcommand{\POWHEG}{{\tt{POWHEG}}}
\newcommand{\PBOX}{{\tt{POWHEG-BOX}}}

\newcommand{\PYTHIA}{{\tt{PYTHIA}}}

\newcommand{\PYTHIAE}{{\tt{PYTHIA8}}}
\newcommand{\HERWIG}{{\tt{HERWIG}}}
\newcommand{\HERWIGS}{{\tt{HERWIG7}}}
\newcommand{\VINCIA}{{\tt{Vincia}}}
\newcommand{\ghhvv}{g_{HHVV}}
\newcommand{\ghvv}{g_{HVV}}
\newcommand{\ghhh}{g_{HHH}}
\newcommand{\lamhhh}{\ghhh}
\newcommand{\klam}{\kappa_\lambda}
\newcommand{\khhvv}{\kappa_{2V}}
\newcommand{\khvv}{\kappa_{V}}

\newcommand{\vbfhh}{VBF~$HH$}
\newcommand{\ptj}{p_{T,j}}
\newcommand{\muc}{\mu_0}
\newcommand{\muf}{\mu_\mathrm{F}}
\newcommand{\mur}{\mu_\mathrm{R}}
\newcommand{\xif}{\xi_\mathrm{F}}
\newcommand{\xir}{\xi_\mathrm{R}}
\newcommand{\pthh}{p_{T,HH}}

\newcommand{\clam}{c_\lambda}
\newcommand{\cv}{c_V}
\newcommand{\cvv}{c_{2V}}

\usepackage{comment}

\usepackage[dvipsnames]{xcolor}

\begin{document}

\pagestyle{SPstyle}
\begin{flushright}\footnotesize{LHCHWG-2025-012, CERN-TH-2025-170, KA-TP-27-2025, P3H-25-062, SI-HEP-2025-18}\end{flushright}
\begin{center}{\Large \textbf{\color{scipostdeepblue}{
        State-of-the-art electroweak Higgs boson pair production in association with two jets at the LHC in the Standard Model and beyond
}}}\end{center}

\begin{center}\textbf{
Jens Braun\textsuperscript{1},
Pia Bredt\textsuperscript{2},
Gudrun Heinrich,\textsuperscript{1},
Marius H\"ofer,\textsuperscript{1},
Barbara J\"ager\textsuperscript{3},
Alexander Karlberg\textsuperscript{4},
Simon Reinhardt\textsuperscript{3}
}\end{center}

\begin{center}
{\bf 1} Institute for Theoretical Physics, Karlsruhe Institute
  of Technology (KIT), 76131 Karlsruhe, Germany
\\
{\bf 2} Department of Physics, University of Siegen, 57068 Siegen, Germany
\\
{\bf 3} Institute for Theoretical Physics, University of Tübingen, Auf der Morgenstelle 14,   72076~Tübingen,
Germany
\\
{\bf 4} CERN, Theoretical Physics Department, 1211, Geneva 23, Switzerland
\\[\baselineskip]
\href{mailto:j.braun@kit.edu}{\small j.braun@kit.edu}, \href{mailto:pia.bredt@uni-siegen.de}{\small pia.bredt@uni-siegen.de}, \href{mailto:gudrun.heinrich@kit.edu}{\small gudrun.heinrich@kit.edu}, 
\href{mailto:marius.hoefer@kit.edu}{\small marius.hoefer@kit.edu},
\href{mailto:jaeger@itp.uni-tuebingen.de}{\small jaeger@itp.uni-tuebingen.de},
\href{mailto:alexander.karlberg@cern.ch}{\small alexander.karlberg@cern.ch},
\href{mailto:simon.reinhardt@uni-tuebingen.de}{\small simon.reinhardt@uni-tuebingen.de}
\end{center}

\section*{\color{scipostdeepblue}{Abstract}}
\textbf{\boldmath{%
We present a systematic comparison of two state-of-the-art tools for the simulation of Higgs boson pair production via vector boson fusion (VBF) as implemented in the Monte-Carlo tools \GOWIZ{} and the \PBOX{}. Cross sections and distributions are provided within the Standard Model and beyond, within scenarios typical for experimental physics analyses, and for a range of energies of relevance to the LHC and its upcoming high luminosity phase. We further perform a detailed study of the so-called VBF approximation, in particular in the presence of anomalous Higgs boson couplings.
%
}}

\vspace{\baselineskip}

\noindent\textcolor{white!90!black}{%
\fbox{\parbox{0.975\linewidth}{%
\textcolor{white!40!black}{\begin{tabular}{lr}%
  \begin{minipage}{0.6\textwidth}%
    {\small Copyright attribution to authors. \newline
    This work is a submission to SciPost Phys. Comm. Rep. \newline
    License information to appear upon publication. \newline
    Publication information to appear upon publication.}
  \end{minipage} & \begin{minipage}{0.4\textwidth}
    {\small Received Date \newline Accepted Date \newline Published Date}%
  \end{minipage}
\end{tabular}}
}}
}\newpage


\vspace{10pt}
\noindent\rule{\textwidth}{1pt}
\tableofcontents
\noindent\rule{\textwidth}{1pt}
\vspace{10pt}


\section{Introduction}
Following the discovery of the Higgs boson by the ATLAS~\cite{ATLAS:2012yve} and CMS~\cite{CMS:2012qbp} Collaborations at the CERN Large Hadron Collider (LHC) in 2012, the particle physics community is now preparing for the upcoming high luminosity phase of the LHC (HL-LHC)~\cite{Apollinari:2017lan}. One of the main objectives of this machine will be to precisely determine the couplings of the Higgs boson to the other particles of the Standard Model (SM) and, crucially, to itself~\cite{Cepeda:2019klc}. One of the key production channels in this endeavour will be the vector boson fusion (VBF) Higgs boson pair production channel, which provides direct access to the Higgs boson self-coupling. Although the cross section of this process is smaller than the one for gluon fusion Higgs boson pair production by an order of magnitude, it provides a very clean experimental environment due to two energetic forward jets that accompany the Higgs boson pair.  

From a theory point of view, VBF Higgs boson pair production in the SM is very well understood. Next-to-leading order (NLO) QCD corrections are sizeable (reaching $\mathcal{O}(30\%)$ in tails of distributions) and have been known in the {\em VBF approximation} (which is defined in detail in Section~\ref{sec:ewhh}) for more than a decade~\cite{Figy:2008zd,Baglio:2012np,Frederix:2014hta}. They are available in the public codes {\tt VBFNLO}~\cite{Arnold:2008rz,Baglio:2024gyp} and {\tt MadGraph5\_aMC@NLO}~\cite{Alwall:2014hca}. 
Inclusive next-to-next-to-leading order (NNLO) QCD corrections in the same approximation (also known as factorizable NNLO corrections), were first calculated in Ref.~\cite{Ling:2014sne}, and have even been extended to next-to-next-to-next-to-leading order (N3LO) in Ref.~\cite{Dreyer:2018qbw}. Beyond the inclusive level, factorizable NNLO corrections are fully known~\cite{Dreyer:2018rfu} whereas non-factorizable corrections are only known in the leading eikonal approximation~\cite{Dreyer:2020urf}. The publicly available code {\tt proVBFHH}~\cite{proVBFHH} contains all NNLO and N3LO corrections.

In Ref.~\cite{Dreyer:2020xaj} the NLO electroweak (EW) corrections were computed and combined with NNLO QCD corrections, and the VBF approximation was studied.

Higgs boson pair production through VBF beyond the SM has been studied extensively in the literature using Effective Field Theories (EFTs)~\cite{Ling:2017teo,Arganda:2018ftn,Araz:2020zyh,Kilian:2018bhs,Dedes:2025oda}, and with the aim of distinguishing linear from non-linear realisations of the Higgs sector~\cite{Gomez-Ambrosio:2022qsi,Gomez-Ambrosio:2022why,Herrero:2022krh,Anisha:2024ljc,Anisha:2024ryj,Domenech:2025gmn}. Crucially none of the studies include NLO QCD corrections.


Recently, two Monte Carlo programs that can provide NLO QCD corrections combined with the leading anomalous coupling contributions to this process have become available, one within the \POWHEG{} framework~\cite{Jager:2025isz}, and another one based on {\tt GoSam+Whizard}~\cite{Braun:2025hvr}. They differ mainly in their application of the VBF approximation. Whereas the \POWHEG{} tool applies the VBF approximation, {\tt GoSam+Whizard} retains all EW diagrams contributing to the $HHjj$ topology. This allows us to perform a detailed study of the VBF approximation also in the presence of both NLO QCD corrections and anomalous couplings. 

 In this paper we therefore carry out a comparative study of the two codes presented in Refs.~\cite{Jager:2025isz,Braun:2025hvr}. In Section~\ref{sec:ewhh} we define the process and the terminology, before identifying the leading anomalous couplings in the context of Higgs Effective Field Theory (HEFT). We briefly describe the technical aspects of the two tools in Section~\ref{sec:mctools}, and provide a range of results in the form of plots and tables of cross section values at $\sqrt{s}=13, 13.6$ and 14\,TeV in Section~\ref{sec:results:vbfapprox}. In addition to predictions in the SM, we show cross sections at four benchmark points in the anomalous coupling parameter space. Section~\ref{sec:results:shower} is dedicated to the assessment of parton-shower effects and in Section~\ref{sec:conclusion} we conclude.

\section{Electroweak Higgs boson pair production in association with two jets}\label{sec:ewhh}
\begin{figure}[t]
    \centering
    \begin{subfigure}[b]{0.33\textwidth}
        \centering
        \includegraphics[scale=1]{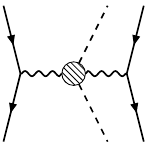}
        \caption{}\label{fig:s-channel}
    \end{subfigure}%
~
    \begin{subfigure}[b]{0.66\textwidth}
        \centering
        \includegraphics[scale=1]{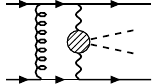}
        \hspace{1cm}  
        \includegraphics[scale=1]{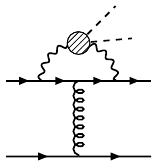}        
        \caption{}\label{fig:non-VBF-NLO}
    \end{subfigure}%
\vspace{0.5cm}
    \begin{subfigure}[b]{0.5\textwidth}
        \centering
        $\includegraphics[scale=1,valign=c]{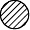} \in
        \left\{
        \includegraphics[scale=1,valign=c]{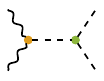},
        \includegraphics[scale=1,valign=c]{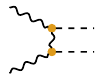},
        \includegraphics[scale=1,valign=c]{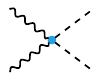}
        \right\}$
        \caption{}\label{fig:VBF_LO_sub}
    \end{subfigure}
    \caption{Examples of diagrams that are not included in the VBF approximation. The shaded blob in diagrams~\ref{fig:s-channel} and~\ref{fig:non-VBF-NLO} can represent any of the sub-diagrams shown in~\ref{fig:VBF_LO_sub}.}\label{fig:nonVBFapprox}
\end{figure}

In this paper we investigate the EW production of two Higgs bosons in association with two jets, i.e. the process $pp\to HHjj$ at $\mathcal{O}\left(\alpha^4\right)$, where $\alpha$ is the EW coupling constant. Neglecting the coupling of the Higgs boson to light quarks, there are two classes of contributions at the Born level to this process: on one hand the $t$-channel and $u$-channel scattering of two light quarks with the exchange of an EW vector boson, which radiates off the Higgs bosons (\emph{VBF process}), and on the other hand the s-channel topology with a $q\bar{q}$ pair annihilating into an EW vector boson with subsequent decay into $q\bar{q}$ again (\emph{Higgs strahlungs process}, see Fig.~\ref{fig:s-channel}). Interferences between $t$-channel and $u$-channel contributions as well as all $s$-channel contributions are known to be suppressed by the event selection cuts used in experiments~\cite{Ciccolini:2007ec,Dreyer:2020xaj}. The one-loop QCD corrections to the process then comprise all possible ways to attach a virtual gluon to the quark lines. In the \emph{VBF approximation}, only the VBF process is considered, and any gluon exchange between the two different quark lines is neglected. When instead including those contributions, we will speak of the \emph{full process}. To suppress contributions from processes that are not of VBF type, typically so-called \emph{VBF cuts} are applied on the invariant mass and the rapidity separation of the tagging jets.
Examples of diagrams not present in the VBF approximation are shown in Fig.~\ref{fig:nonVBFapprox}. Note that since in the full process all $s$-, $t$- and $u$-channel crossings and their interferences are taken into account, the contribution of diagrams such as in Fig.~\ref{fig:non-VBF-NLO} does not vanish. This is different when only the $t$-channel and the $u$-channel contributions are considered separately, because of colour conservation. See Ref.~\cite{Braun:2025hvr} for more details.

\subsection{Higgs Effective Field Theory (HEFT)}

While a full description of the framework of Higgs Effective Field Theory, also called HEFT or {\em Electroweak Chiral Lagrangian}~\cite{Feruglio:1992wf,Bagger:1993zf,Burgess:1999ha,Grinstein:2007iv,Alonso:2012px,Buchalla:2012qq,Buchalla:2013rka}, is beyond the scope of this paper, we briefly introduce some basic features here, such that we can identify the set of leading operators contributing to Higgs boson pair production in VBF.

In contrast to the Standard Model Effective Field Theory (SMEFT)~\cite{Buchmuller:1985jz,Grzadkowski:2010es,Brivio:2017vri,Isidori:2023pyp}, the Higgs field in the HEFT is not assumed to be part of a doublet of the electroweak $SU(2)$ symmetry. As a consequence, the Higgs field is treated independently from the Goldstone bosons and obeys fewer constraints than in the SMEFT.
The EFT expansion is based on the so-called chiral dimension $d_\chi$, which has a direct correspondence to the loop order $L$ of a given operator, $d_\chi=2L+2$. The Lagrangian that is LO in the HEFT expansion collects all terms of chiral dimension $d_\chi=2$, therefore we call it $\mathcal{L}_2$. The NLO (in the HEFT power counting) Lagrangian has chiral dimension $d_\chi=4$, corresponding to loop order $L=1$, and is denoted by $\mathcal{L}_4$.
The main features of $\mathcal{L}_2$ are that the Higgs potential $V(h)$ can contain arbitrary powers of the Higgs field $h$, since $h$ has chiral dimension zero,
\begin{align}
    V(\eta) &= v^4\sum_{n=2}^\infty V_n\eta^n\quad ; \quad \eta=\frac{h}{v}\;,
\end{align}
where $v$ is the scale of electroweak symmetry breaking and $V_n$ are $\mathcal{O}(1)$ coefficients. The term in the Lagrangian containing the covariant derivatives contains the so-called {\em flare function},  
\begin{align}
    F(\eta) &= 1 + \sum_{n=1}^\infty F_n\eta^n\;,
\end{align}
where $F_n$ are $\mathcal{O}(1)$ coefficients, such that the part of $\mathcal{L}_2$ containing the interactions of the Higgs boson with the gauge and Goldstone bosons reads
\begin{align}
    \mathcal{L}^{\rm{gauge-Higgs}}_2 =&  \frac{v^2}{2}\partial_\mu\eta\,\partial^\mu\eta - V(\eta) + \frac{v^2}{4}\left\langle (D_\mu U)^\dagger (D^\mu U)\right\rangle F(\eta)\label{eq:HEFT_partialL2}\,,
\end{align}
with \begin{align}
\label{eq:cov-deriv}
    D_\mu U &= \partial_\mu U + igW_\mu U - ig'B_\mu Ut^3\,, 
\end{align}
and $U=\exp(2i\varphi/v)$, where $\varphi=\varphi^\alpha t^\alpha$ with $\alpha=1,2,3$ are the Goldstone boson fields and the $t^\alpha$ are the generators of the $SU(2)_L$ gauge group. $W_\mu=W_\mu^\alpha t^\alpha$ and $B_\mu$ denote the $SU(2)_L$ and $U(1)_Y$ gauge fields, respectively, with associated gauge couplings $g$ and $g'$. 
The gauge-fermion interactions remain SM-like at LO in the HEFT expansion.
We omit contributions to $\mathcal{L}_2$ that violate custodial symmetry because they would generate a tree-level contribution to the electroweak $\rho$-parameter, which would have been noticed experimentally~\cite{Veltman:1977kh,ParticleDataGroup:2024cfk}.

Finally, we note that the process, within the VBF approximation, in the context of the SMEFT could in principle be studied using the MadGraph5\_aMC@NLO~\cite{Frederix:2014hta} framework in conjunction with SMEFT@NLO~\cite{Degrande:2020evl}. However, to our knowledge such a study has never been published.

\subsection{Effective HEFT Lagrangian for Higgs boson pair production in VBF}

Focusing on Higgs boson pair production in VBF, the leading anomalous interactions resulting from $\mathcal{L}_2$ can be parametrised as follows: 
\begin{align}
    \mathcal{L}_\mathrm{eff} &\supset \left(2c_V\frac{h}{v}+c_{2V}\frac{h^2}{v^2}\right)\left(m_W^2W_\mu^+W^{-\mu}+\frac{1}{2}m_Z^2Z_\mu Z^\mu\right) -c_\lambda\frac{m_h^2}{2v}h^3\,,\label{eq:LO_eff}
\end{align}
where $W_\mu^\pm = (W_\mu^1\mp W_\mu^2)/\sqrt{2}$ and $Z_\mu=\cos\theta_WW_\mu^3-\sin\theta_WB_\mu$, with $\theta_W$ the Weinberg angle just as in the SM. Note that no new structures arise which are not present in the SM and that for $c_V=c_{2V}=c_\lambda=1$ we recover the SM. Operators from $\mathcal{L}_4$ are not included in this analysis, which we motivate by the fact that we focus on the NLO QCD corrections. 
In principle, performing a calculation beyond tree level implies that also operators with chiral dimension $d_\chi=4$, i.e.\ one-loop order, have to be considered, because
tree-level amplitudes with a single vertex insertion from $\mathcal{L}_4$ are of the same order as genuine one-loop amplitudes constructed from interactions contained in $\mathcal{L}_2$ only.
However, here we restrict ourselves to NLO in QCD, which means that the one-loop corrections are ${\cal O}(\alpha_s)$ in the strong coupling $\alpha_s$ relative to the Born contributions. These Born contributions are ${\cal O}(\alpha^4)$ in the EW coupling $\alpha$ for Higgs boson pair production in VBF. Thus, in our NLO QCD calculation, we include all coupling powers at ${\cal O}(\alpha^4\alpha_s)$. 
Although the HEFT operators do not exhibit explicit powers in either the electroweak or strong coupling, one can still make a formal assignment based on their possible couplings to SM particles in a renormalisable UV completion. Furthermore, from power counting arguments, it follows that each Higgs boson introduces a factor of $1/v\propto \sqrt{4\pi\alpha}$. Therefore, if SM NLO EW  corrections were included in the calculation, this would require us to also consider operators from $\mathcal{L}_4$, to be consistent in the HEFT power counting.

\subsection{The {\em kappa framework}}
The effective Lagrangian from Eq.~(\ref{eq:LO_eff}) coincides with the settings of the so-called {\em kappa framework}~\cite{Hagiwara:1993ck,LHCHiggsCrossSectionWorkingGroup:2012nn},
a simple prescription to parametrise deviations from the SM. In this framework, a coupling modifier $\kappa_i$ is defined as the ratio of a coupling $g_i$ to the corresponding SM coupling $g_i^{SM}$.  In particular, the couplings of the Higgs boson entering the VBF-induced $HHjj$ production are given by
\begin{align} 
    \ghvv &= \khvv \cdot \ghvv^\mathrm{SM}\,, &
    \ghhvv &= \khhvv \cdot \ghhvv^\mathrm{SM}\,, &    
    \lamhhh &= \klam \cdot \lamhhh^\mathrm{SM}\,.
\end{align} 
Here, $\ghvv$ denotes the trilinear coupling of a Higgs boson to two massive weak bosons, $\ghhvv$ the quartic coupling between two Higgs bosons and two massive weak bosons and $\lamhhh$ the trilinear Higgs coupling.
In the following, we will express the coupling factors $\kappa_i$ in terms of the $c_i$ introduced in Eq.~\eqref{eq:LO_eff}, $\kappa_i=c_i$, and denote points in the parameter space of anomalous couplings in the format $\left[c_{\lambda},c_V,c_{2V}\right]$.
 
\section{Description of the Monte Carlo programs}\label{sec:mctools}
\subsection{\GOSAM+\WHIZARD}
\GOSAM~\cite{Cullen:2011ac,GoSam:2014iqq,Braun:2025afl} is a program package to produce and evaluate one-loop amplitudes in an automated way. It generates algebraic expressions for Feynman diagrams with the help of {\sc qgraf}~\cite{Nogueira:1991ex} and {\sc form}~\cite{Vermaseren:2000nd,Kuipers:2012rf}. Subsequently, integral reduction is performed using the tool {\sc Ninja}~\cite{vanDeurzen:2013saa,Peraro:2014cba} and the master integrals are evaluated with {\sc OneLoop}~\cite{vanHameren:2010cp}.
It can import arbitrary model files written in the UFO format~\cite{Degrande:2011ua,Darme:2023jdn}. The new version \GOSAM-3.0~\cite{Braun:2025afl} has additional EFT functionalities.
\WHIZARD~\cite{Kilian:2007gr,Moretti:2001zz} is a multi-purpose Monte Carlo program providing phase space integration and event generation for hadron and lepton collider processes. It includes an automation of NLO SM corrections \cite{ChokoufeNejad:2016qux, Stienemeier:2021cse,Bredt:2022zpn} relying on the FKS subtraction scheme \cite{Frixione:1995ms}.
The NLO events can be matched to a parton shower through an implementation of the POWHEG NLO matching scheme~\cite{Nason:2004rx,Frixione:2007vw} in \WHIZARD~\cite{ChokoufeNejad:2015kpc,Stienemeier:2022wmy}.

The amplitudes at the tree and one-loop level for the full EW $HHjj$ process are computed by \GOSAM{} and passed to \WHIZARD{} via its interface based on the Binoth Les Houches Accord (BLHA) standard~\cite{Binoth:2010xt,Alioli:2013nda}.

The combined setup including EFT functionalities was first presented in \cite{Braun:2025hvr} and is available at \url{https://github.com/Jens-Braun/VBF_HH_HEFT}.
 
\subsection{\PBOX}
The \PBOX{}\cite{Alioli:2010xd} constitutes a general framework for the matching of fixed-order perturbative calculations with parton-shower (PS) programs using the \POWHEG{} formalism~\cite{Nason:2004rx,Frixione:2007vw}. An implementation of the VBF-induced Higgs boson pair production process has been presented in \cite{Jager:2025isz} and is publicly available from the webpage of the \PBOX{} project at \url{https://powhegbox.mib.infn.it/}. This tool allows for the calculation of NLO-QCD corrections within the VBF approximation and their combination with multi-purpose Monte Carlo generators such as \HERWIG{}~\cite{Bewick:2023tfi} or \PYTHIA{}~\cite{bierlich2022comprehensiveguidephysicsusage}. At fixed order, both with and without anomalous couplings, it has been validated against the \PROVBFHH{} program~\cite{Dreyer:2018rfu,Dreyer:2018qbw,Dreyer:2020urf}. In the following, we will refer to NLO-QCD predictions matched with a PS as NLO+PS results.

\section{Results}\label{sec:results}
We perform numerical studies for the three proton-proton centre-of-mass energies $\sqrt{s}=13, $ 13.6 and $14\,$TeV. For the discussion of differential distributions, we focus on the latter. We use the PDF4LHC21\_40 set of parton distribution functions (PDFs)~\cite{PDF4LHCWorkingGroup:2022cjn}. Jets are reconstructed with the anti-$k_t$~algorithm~\cite{Cacciari:2008gp} using a jet radius parameter of $R=0.4$. 
As electroweak input parameters we choose the fine structure constant $\alpha\left(M_Z\right)=1/127.9$, the Fermi constant, $G_F=1.16637\times 10^{-5}$~GeV$^{-2}$, and the mass of the $Z$~boson, $M_Z=91.1876$~GeV. The widths of the $Z$ and $W$~bosons are set to $\Gamma_Z=2.4952$~GeV and $\Gamma_W=2.085$~GeV, respectively~\cite{ParticleDataGroup:2024cfk}. For the Higgs boson, we use a mass of $M_H=125.09$~GeV and a width of $\Gamma_H=3.7$~MeV. 
The factorisation and renormalisation scales, $\muf=\xif \muc$ and $\mur=\xir\muc$, are computed from the mass of the Higgs boson and the transverse momentum of the Higgs boson pair system, $\pthh$, according to~\cite{Dreyer:2018rfu},  
\begin{equation}
    \muc = \sqrt{\frac{M_H}{2}\sqrt{\frac{M_H^2}{4}+\pthh^2}}\,.
\end{equation} 
For the assessment of scale uncertainties, we perform a seven-point variation of the scale parameters $\xif,\xir$ in the range 0.5 to 2. 

Our default selection cuts are designed to favour VBF-induced $HHjj$ production. To that end we require the presence of at least two jets with transverse momenta and rapidities in the ranges
\begin{equation}
\label{eq:jcuts-def}
    \ptj> 25~\mathrm{GeV}\,, \quad
    |y_j|<4.5\,.
\end{equation}
The two hardest (highest transverse momentum) jets that meet these criteria are called {\em tagging jets}. On the tagging jets, we additionally impose requirements on their invariant mass and pseudorapidity separation, 
\begin{equation}
\label{eq:jjcuts-def}
    m_{jj}^\mathrm{tag}> 600~\mathrm{GeV}\,, \quad
    \Delta \eta_{jj}^\mathrm{tag}=|\eta_{j_1}^\mathrm{tag}-\eta_{j_2}^\mathrm{tag}|>4.5\,.
\end{equation}
In the following we will refer to the cuts of Eqs.~\eqref{eq:jcuts-def}--\eqref{eq:jjcuts-def} as {\em VBF cut setup}, in contrast to an {\em inclusive cut setup}, where we only impose the cuts of Eq.~\eqref{eq:jcuts-def}. 

Whenever we consider subleading jets, we require these to exhibit transverse momenta and rapidities in the ranges  
\begin{equation}
\label{eq:jets-sub}
    p_{T,j_{sub}}> 25~\mathrm{GeV}\,, \quad
    |y_{j_{sub}}|<4.5\,.
\end{equation}

\subsection{Assessment of the VBF approximation}\label{sec:results:vbfapprox}
The various tools presented in this work are designed to provide predictions for the \vbfhh{} process within the SM and in the presence of anomalous Higgs couplings, including different subsets of perturbative contributions. To explore whether they are equivalently suited for the description of experimentally relevant scenarios, we have performed a detailed comparison of cross sections and characteristic differential distributions for different setups.  

In Tabs.~\ref{tab:xsec-inc} and \ref{tab:xsec-vbf} we present results for cross sections of the EW~$HHjj$ production process at LO and NLO-QCD accuracy obtained with the  \GOSAM+\WHIZARD{} and \PBOX{} programs for several values of the LHC centre-of-mass energy and the two cut scenarios introduced above. 
The benchmark points in the space of anomalous couplings shown in Tabs.~\ref{tab:xsec-inc} and \ref{tab:xsec-vbf} have been chosen such that they lead to characteristic shape changes in differential cross sections such as the invariant mass and rapidity of the di-Higgs system, see Ref.~\cite{Braun:2025hvr} for more details.

Because the \GOSAM+\WHIZARD{} implementation includes the full EW~$HHjj$ matrix elements, while the \PBOX{} implementation resorts to the VBF approximation, see section~\ref{sec:ewhh}, differences between predictions obtained with the two programs are expected to be larger for inclusive setups than in VBF-specific phase-space regions.

Indeed, for the {\bf inclusive cut} setup, we observe in Tab.~\ref{tab:xsec-inc} differences of almost 46\% (30\%) for the NLO (LO) cross sections for the SM predictions, and in the presence of anomalous couplings the differences can even exceed 50\%.
In contrast, for the {\bf VBF cut} setup, the differences are at the one percent level both at LO and NLO, also in the presence of anomalous couplings, as can be seen in Tab.~\ref{tab:xsec-vbf}. The dependence of the difference on the centre-of-mass energy is only marginal.

Let us emphasise that in typical VBF analyses where cuts are applied that enhance the relative contribution of VBF topologies to the full $HHjj$ production process, the use of tools relying on the VBF approximation is fully satisfactory. If, however, more inclusive selection cuts are applied, contributions neglected within the VBF approximation can become highly relevant, as we will investigate further in the following section by means of various differential distributions in the SM case.

\begin{table}[pt!]
\begin{center}
\begin{tabular}[t]{|c|ll|ll|ll|}
\toprule
$\sqrt{s} = 13\, \mathrm{TeV}$ & \multicolumn{2}{c|}{\GOSAM+\WHIZARD} & \multicolumn{2}{c|}{\PBOX} & \multicolumn{2}{c|}{Ratio} \\
\midrule
$\left[ c_{\lambda}, c_{V}, c_{2V} \right] $ & {$\sigma_{\mathrm{NLO}}$ [fb]} & $K$ & {$\sigma_{\mathrm{NLO}}$ [fb]} & $K$ & LO & NLO\\
\midrule
$\left[1.0,1.0,1.0\right]$ & $1.387(10)$ & $1.09$ & $0.951(4)$ & $0.97$ & $1.295(3)$ & $1.458(12)$\\
$\left[-1.0,0.9,1.5\right]$ & $2.644(11)$ & $1.02$ & $2.381(7)$ & $0.99$ & $1.079(1)$ & $1.110(6)$\\
$\left[1.0,0.9,1.0\right]$ & $0.888(5)$ & $1.14$ & $0.508(2)$ & $0.97$ & $1.500(2)$ & $1.748(12)$\\
$\left[1.0,1.0,0.5\right]$ & $5.436(26)$ & $1.00$ & $5.168(9)$ & $0.98$ & $1.029(1)$ & $1.052(5)$\\
$\left[4.0,0.95,0.5\right]$ & $4.465(20)$ & $1.05$ & $3.459(7)$ & $0.98$ & $1.194(1)$ & $1.291(6)$\\
\midrule
$\sqrt{s} = 13.6\, \mathrm{TeV}$ & \multicolumn{2}{c|}{\GOSAM+\WHIZARD} & \multicolumn{2}{c|}{\PBOX} & \multicolumn{2}{c|}{Ratio} \\
\midrule
$\left[ c_{\lambda}, c_{V}, c_{2V} \right] $ & {$\sigma_{\mathrm{NLO}}$ [fb]} & $K$ & {$\sigma_{\mathrm{NLO}}$ [fb]} & $K$ & LO & NLO\\
\midrule
$\left[1.0,1.0,1.0\right]$ & $1.509(10)$ & $1.08$ & $1.055(6)$ & $0.97$ & $1.284(3)$ & $1.430(12)$\\
$\left[-1.0,0.9,1.5\right]$ & $3.021(12)$ & $1.02$ & $2.711(7)$ & $0.99$ & $1.077(1)$ & $1.114(5)$\\
$\left[1.0,0.9,1.0\right]$ & $0.988(5)$ & $1.14$ & $0.569(2)$ & $0.97$ & $1.481(2)$ & $1.736(11)$\\
$\left[1.0,1.0,0.5\right]$ & $6.009(26)$ & $0.99$ & $5.778(10)$ & $0.98$ & $1.028(1)$ & $1.040(5)$\\
$\left[4.0,0.95,0.5\right]$ & $4.982(24)$ & $1.07$ & $3.838(8)$ & $0.98$ & $1.188(2)$ & $1.298(7)$\\
\midrule
$\sqrt{s} = 14\, \mathrm{TeV}$ & \multicolumn{2}{c|}{\GOSAM+\WHIZARD} & \multicolumn{2}{c|}{\PBOX} & \multicolumn{2}{c|}{Ratio} \\
\midrule
$\left[ c_{\lambda}, c_{V}, c_{2V} \right] $ & {$\sigma_{\mathrm{NLO}}$ [fb]} & $K$ & {$\sigma_{\mathrm{NLO}}$ [fb]} & $K$ & LO & NLO\\
\midrule
$\left[1.0,1.0,1.0\right]$ & $1.587(11)$ & $1.07$ & $1.128(5)$ & $0.97$ & $1.277(3)$ & $1.407(12)$\\
$\left[-1.0,0.9,1.5\right]$ & $3.270(14)$ & $1.02$ & $2.942(7)$ & $0.99$ & $1.073(1)$ & $1.111(5)$\\
$\left[1.0,0.9,1.0\right]$ & $1.053(5)$ & $1.14$ & $0.614(2)$ & $0.98$ & $1.468(3)$ & $1.715(10)$\\
$\left[1.0,1.0,0.5\right]$ & $6.506(29)$ & $1.00$ & $6.208(11)$ & $0.98$ & $1.028(1)$ & $1.048(5)$\\
$\left[4.0,0.95,0.5\right]$ & $5.228(25)$ & $1.05$ & $4.104(9)$ & $0.98$ & $1.183(1)$ & $1.274(7)$\\
\bottomrule
\end{tabular}
\caption{\label{tab:xsec-inc} Cross sections for the EW $HHjj$ production process at the LHC within the {\bf inclusive cut} setup at NLO-QCD accuracy as obtained with the \GOWIZ{} and \PBOX{} programs for a selection of centre-of-mass~energies, for the SM $\left(\left[ 1.0,1.0,1.0 \right]\right)$ and four benchmark points in the space of anomalous couplings. The numbers in parenthesis denote Monte Carlo integration uncertainties. The uncertainties of the $K$-factors (NLO/LO) are at permille level, i.~e. one order of magnitude below the last $K$-factors' digit. The last two columns show the ratio \GOWIZ{}/\PBOX{} at LO and NLO, respectively.}
\end{center}
\end{table}

\begin{table}[pt!]
\begin{center}
\begin{tabular}[t]{|c|ll|ll|ll|}
\toprule
$\sqrt{s} = 13\, \mathrm{TeV}$ & \multicolumn{2}{c|}{\GOSAM+\WHIZARD} & \multicolumn{2}{c|}{\PBOX} & \multicolumn{2}{c|}{Ratio} \\
\midrule
$\left[ c_{\lambda}, c_{V}, c_{2V} \right] $ & {$\sigma_{\mathrm{NLO}}$ [fb]} & $K$ & {$\sigma_{\mathrm{NLO}}$ [fb]} & $K$ & LO & NLO\\
\midrule
$\left[1.0,1.0,1.0\right]$ & $0.588(5)$ & $0.92$ & $0.585(2)$ & $0.91$ & $0.991(2)$ & $1.005(9)$\\
$\left[-1.0,0.9,1.5\right]$ & $1.614(8)$ & $0.93$ & $1.608(3)$ & $0.92$ & $0.998(1)$ & $1.004(5)$\\
$\left[1.0,0.9,1.0\right]$ & $0.284(2)$ & $0.93$ & $0.2862(9)$ & $0.92$ & $0.983(2)$ & $0.992(8)$\\
$\left[1.0,1.0,0.5\right]$ & $3.416(18)$ & $0.92$ & $3.395(6)$ & $0.92$ & $1.000(1)$ & $1.006(6)$\\
$\left[4.0,0.95,0.5\right]$ & $1.833(13)$ & $0.93$ & $1.806(5)$ & $0.92$ & $0.999(1)$ & $1.015(8)$\\
\midrule
$\sqrt{s} = 13.6\, \mathrm{TeV}$ & \multicolumn{2}{c|}{\GOSAM+\WHIZARD} & \multicolumn{2}{c|}{\PBOX} & \multicolumn{2}{c|}{Ratio} \\
\midrule
$\left[ c_{\lambda}, c_{V}, c_{2V} \right] $ & {$\sigma_{\mathrm{NLO}}$ [fb]} & $K$ & {$\sigma_{\mathrm{NLO}}$ [fb]} & $K$ & LO & NLO\\
\midrule
$\left[1.0,1.0,1.0\right]$ & $0.660(6)$ & $0.93$ & $0.654(2)$ & $0.91$ & $0.992(2)$ & $1.009(10)$\\
$\left[-1.0,0.9,1.5\right]$ & $1.846(10)$ & $0.92$ & $1.850(4)$ & $0.92$ & $0.998(1)$ & $0.998(6)$\\
$\left[1.0,0.9,1.0\right]$ & $0.320(3)$ & $0.92$ & $0.3253(9)$ & $0.92$ & $0.986(2)$ & $0.984(10)$\\
$\left[1.0,1.0,0.5\right]$ & $3.783(19)$ & $0.90$ & $3.826(7)$ & $0.91$ & $1.002(1)$ & $0.989(5)$\\
$\left[4.0,0.95,0.5\right]$ & $2.023(15)$ & $0.91$ & $2.034(5)$ & $0.92$ & $1.000(2)$ & $0.995(8)$\\
\midrule
$\sqrt{s} = 14\, \mathrm{TeV}$ & \multicolumn{2}{c|}{\GOSAM+\WHIZARD} & \multicolumn{2}{c|}{\PBOX} & \multicolumn{2}{c|}{Ratio} \\
\midrule
$\left[ c_{\lambda}, c_{V}, c_{2V} \right] $ & {$\sigma_{\mathrm{NLO}}$ [fb]} & $K$ & {$\sigma_{\mathrm{NLO}}$ [fb]} & $K$ & LO & NLO\\
\midrule
$\left[1.0,1.0,1.0\right]$ & $0.725(8)$ & $0.95$ & $0.704(2)$ & $0.91$ & $0.991(2)$ & $1.030(12)$\\
$\left[-1.0,0.9,1.5\right]$ & $2.042(11)$ & $0.93$ & $2.022(4)$ & $0.91$ & $0.996(1)$ & $1.010(6)$\\
$\left[1.0,0.9,1.0\right]$ & $0.347(3)$ & $0.92$ & $0.353(1)$ & $0.92$ & $0.981(2)$ & $0.983(9)$\\
$\left[1.0,1.0,0.5\right]$ & $4.179(22)$ & $0.92$ & $4.132(8)$ & $0.91$ & $1.002(1)$ & $1.011(6)$\\
$\left[4.0,0.95,0.5\right]$ & $2.162(16)$ & $0.90$ & $2.194(6)$ & $0.92$ & $1.002(2)$ & $0.985(8)$\\
\bottomrule
\end{tabular}
\caption{
\label{tab:xsec-vbf} Cross sections for the EW $HHjj$ production process at the LHC within the {\bf VBF cut} setup at NLO-QCD accuracy as obtained with the \GOWIZ{} and \PBOX{} programs for a selection of centre-of-mass~energies, for the SM $\left(\left[ 1.0,1.0,1.0 \right]\right)$ and four benchmark points in the space of anomalous couplings. The numbers in parenthesis denote Monte Carlo integration uncertainties. The uncertainties of the $K$-factors (NLO/LO) are at permille level, i.~e. one order of magnitude below the last $K$-factors' digit. The last two columns show the ratio \GOWIZ{}/\PBOX{} at LO and NLO, respectively.
}
\end{center}
\end{table}

\subsubsection{Standard Model}

\begin{figure}[pt!]
\centering
\begin{subfigure}[l]{0.49\textwidth}
    \centering
    \includegraphics[width=\textwidth,page=2]{Plots-SM_s.pdf} \vspace{0.25cm}
\end{subfigure}
\begin{subfigure}[r]{0.49\textwidth}
    \centering
    \includegraphics[width=\textwidth,page=1]{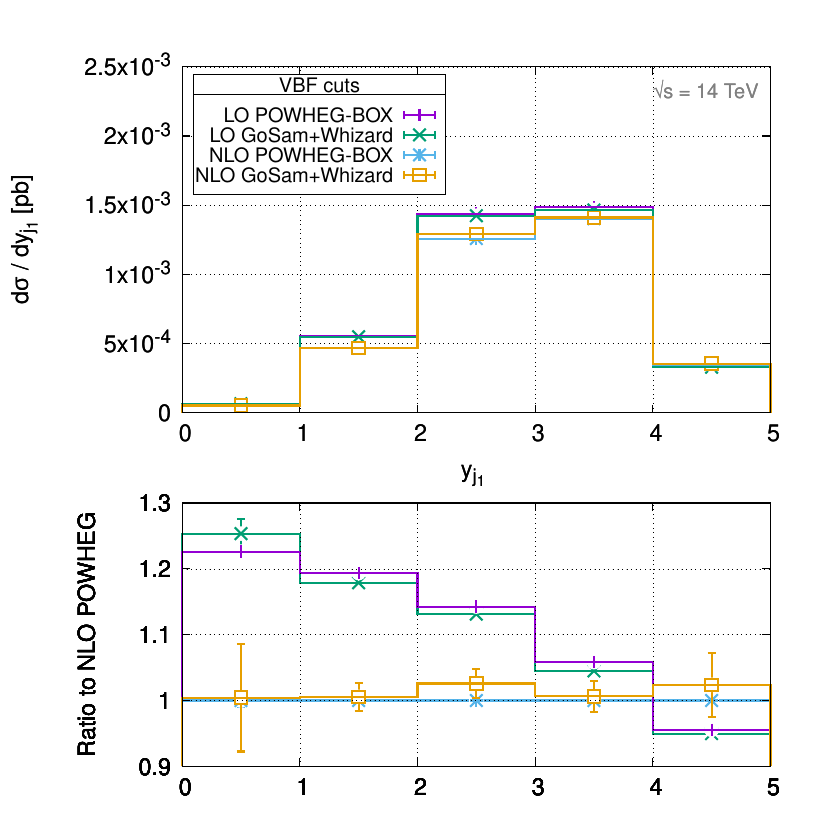} \vspace{0.25cm}
\end{subfigure}
\begin{subfigure}[l]{0.49\textwidth}
    \centering
    \includegraphics[width=\textwidth,page=4]{Plots-SM_s.pdf}
\end{subfigure}
\begin{subfigure}[r]{0.49\textwidth}
    \centering
    \includegraphics[width=\textwidth,page=3]{Plots-SM_s.pdf}
\end{subfigure}
\caption{\label{fig:SM} 
Distributions of the rapidity of the hardest tagging jet (top) and transverse momentum of the hardest Higgs boson (bottom) for the \vbfhh{} process within the {\bf inclusive cut} setup (left)  and the {\bf VBF cut} setup (right)
as obtained with the \PBOX~(LO: purple, NLO: blue) and \GOSAM+\WHIZARD~(LO: green, NLO: orange), and their ratios to the \PBOX~results  at NLO (lower panels).}
\end{figure}

As examples, in Fig.~\ref{fig:SM} we show distributions at $\sqrt{s}=14\,$TeV for the rapidity of the hardest tagging jet and for the transverse momentum of the hardest Higgs boson, respectively, illustrating the difference between the \PBOX~ and \GOSAM+\WHIZARD~predictions for the {\bf inclusive cut} setup and the {\bf VBF cut} setup, in the SM. 

We see clear differences for the {\bf inclusive cut} setup, with factors of up to $2.5$ ($3.1$) at LO (NLO) in the low-rapidity region and up to $1.7$ ($2.2$) at LO (NLO) in the high $p_{T,H_h}$ regime, with $H_h$ being the hardest of the two Higgs bosons. These enhancements of differential cross sections with respect to the VBF approximation can be attributed to contributions of Higgs-strahlung diagrams which are pronounced especially in these regimes. Here the contribution of QCD corrections also dominantly comes from diagrams of Higgs-strahlung type which are not included in the VBF approximation results. This relation between the two different diagram selections can be also inferred by the total cross sections in Tab.~\ref{tab:xsec-inc}, where the ratios in the last two columns are at about $1.3$ ($1.4$) at LO (NLO) and the $K$-factors of \GOSAM+\WHIZARD~ exceed those of \PBOX~ by about $10\%$.

For the {\bf VBF cut} setup we only get differences of about 1\% at LO for $y_{j_1}$ and fairly good agreement at NLO in the range of the uncertainties, establishing the validity of the VBF approximation when VBF cuts are applied. Differences between the full process and the VBF approximation in the {\bf VBF cut} setup are most pronounced in the tail of the $p_{T,H_h}$ distribution, where the full process lies up to 7\% below the VBF approximation at LO and up to 9\% at NLO, fully consistent with Ref.~\cite{Dreyer:2020xaj}. Again this suppression in the tail can be attributed to increasing contributions of diagrams of Higgs-strahlung type. In contrary to the {\bf inclusive setup}, the $K$-factor however can be observed to be approximately of the same size for both tools considering the complete range of $y_{j_1}$ and $p_{T,H_h}$. The differential cross section is reduced by up to $19\%$ in the low $y_{j_1}$ region and $13\%$ for low $p_{T,H_h}$, for both tools, which is also fully consistent with Ref.~\cite{Dreyer:2020xaj} within the {\bf VBF cut} setup. This behaviour is also reflected in the $K$-factors in Tab.~\ref{tab:xsec-vbf}. The QCD corrections are thus found to contribute dominantly for diagrams of the pure VBF process.

We can conclude that in the \textbf{VBF cut setup} the effect of the NLO corrections on the distribution is covered well in the VBF approximation, i.~e. $\mathcal{O}(\alpha_s)$ corrections to the full process beyond the approximation are much suppressed. Due to this observation and the fact that the ratios of total cross section in the last columns of Tab.~\ref{tab:xsec-vbf} deviate from one at most by a few percent, for both LO and NLO, for all benchmark points of anomalous couplings, we only show LO results for the comparison beyond the SM in the next subsection.

\subsubsection{Anomalous couplings}

\begin{figure}[pt!]
\centering
\begin{subfigure}[l]{0.49\textwidth}
    \centering
    \includegraphics[width=\textwidth,page=1]{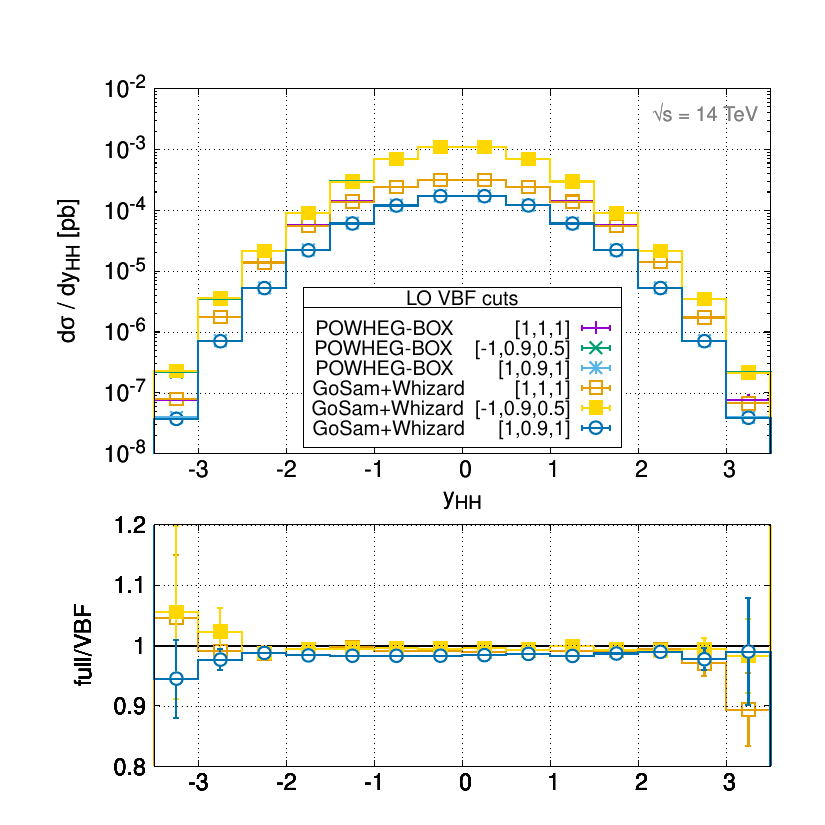} \vspace{0.25cm}
\end{subfigure}
\begin{subfigure}[r]{0.49\textwidth}
    \centering
    \includegraphics[width=\textwidth,page=3]{Plot-LO-anom_s.pdf} \vspace{0.25cm}
\end{subfigure}
\begin{subfigure}[l]{0.49\textwidth}
    \centering
    \includegraphics[width=\textwidth,page=2]{Plot-LO-anom_s.pdf} \vspace{0.25cm}
\end{subfigure}
\begin{subfigure}[r]{0.49\textwidth}
    \centering
    \includegraphics[width=\textwidth,page=4]{Plot-LO-anom_s.pdf} \vspace{0.25cm}
\end{subfigure}
\caption{\label{fig:LO-anom} 
Rapidity (top) and invariant mass (bottom) distributions of the di-Higgs system for the \vbfhh{} process within the {\bf VBF cut} setup at LO as obtained with the \PBOX~and \GOSAM+\WHIZARD~ (upper panels), and the ratios of the full process (\GOSAM+\WHIZARD) to the VBF approximation (\PBOX) for each set of anomalous couplings (lower panels). The values chosen for the anomalous Higgs couplings are quoted in the format $\left[\clam,\cv,\cvv\right]$ for each curve in the legend. In addition to the SM results, denoted by $[1,1,1]$, predictions for the values $[-1,0.9,0.5]$ and $[1,0.9,1]$ are shown on the left, and for the values $[1,1,0.5]$ and $[4,0.95,0.5]$ on the right.}
\end{figure}

In Fig.~\ref{fig:LO-anom} we display LO predictions at $\sqrt{s}=14\,$TeV for the rapidity $y_{HH}$ and invariant mass $m_{HH}$ of the di-Higgs system, respectively. 
We show results obtained with the \PBOX~and \GOSAM+\WHIZARD~ for the SM and different values for the anomalous coupling factors as listed in Tab.~\ref{tab:xsec-vbf} within the {\bf VBF cut} setup. We do not display results with anomalous couplings in the {\bf inclusive cut} setup, because the differences between the full process compared to the VBF approximation exceed the differences due to the inclusion of anomalous couplings in the inclusive setup. We therefore only consider the {\bf VBF cut} setup for this discussion. The anomalous couplings we have chosen are representative for characteristic shapes in differential cross sections while still being within experimental constraints~\cite{Braun:2025hvr,ATLAS:2024ish,CMS:2024awa}.

For the {\bf VBF cut} setup, we notice very good agreement between the two implementations, not only for the SM but also in the presence of anomalous couplings.
The anomalous couplings lead to a significant change in the shape of the distributions compared to the SM, especially in the  di-Higgs invariant mass distributions, while the rapidity distributions rather show overall shifts.

From the upper row of Fig.~\ref{fig:LO-anom}, we see that the VBF approximation only leads to differences at large rapidity of the di-Higgs system, which are not significant given the uncertainties, however.

For the di-Higgs invariant mass distribution, shown in the lower row of Fig.~\ref{fig:LO-anom}, the differences between the full process compared to the VBF approximation are in the $2\%$ range, where the full result tends to be slightly lower than the VBF approximation. However, changes in the value of $c_{2V}$ (see for example point [1,1,0.5]) seem to shift the {\em full/VBF} ratio upward by a few percent compared to the SM ratio.

\subsection{Assessment of the impact of the parton shower}\label{sec:results:shower}
In order to assess the impact of parton-shower radiation on fixed-order results for the \vbfhh{} process, we employ a selection of different multi-purpose Monte Carlo generators. 
For our default {\bf VBF cut} setup with the cuts of Eqs.~\eqref{eq:jcuts-def}--\eqref{eq:jjcuts-def}, we provide predictions obtained with our \PBOX{} implementation matching the NLO-QCD calculation with \HERWIGS{} (version~7.3.0)~\cite{Bewick:2023tfi} and \PYTHIAE{} (version 8.312)~\cite{bierlich2022comprehensiveguidephysicsusage} using the {\tt Monash} tune~\cite{Skands_2014}. For \PYTHIAE{} we consider both a local dipole shower and the \VINCIA{} antenna shower~\cite{Fischer:2016vfv}. We note that using the global default shower of \PYTHIAE{} is not recommended for the simulation of VBF~processes~\cite{Jager:2020hkz,Hoche:2021mkv}, being unsuitable to account for the genuine radiation pattern. 
Underlying event and hadronisation are switched off unless specified otherwise. 

\begin{figure}[pt!]
\centering
\includegraphics[width=0.49\textwidth,page=1]{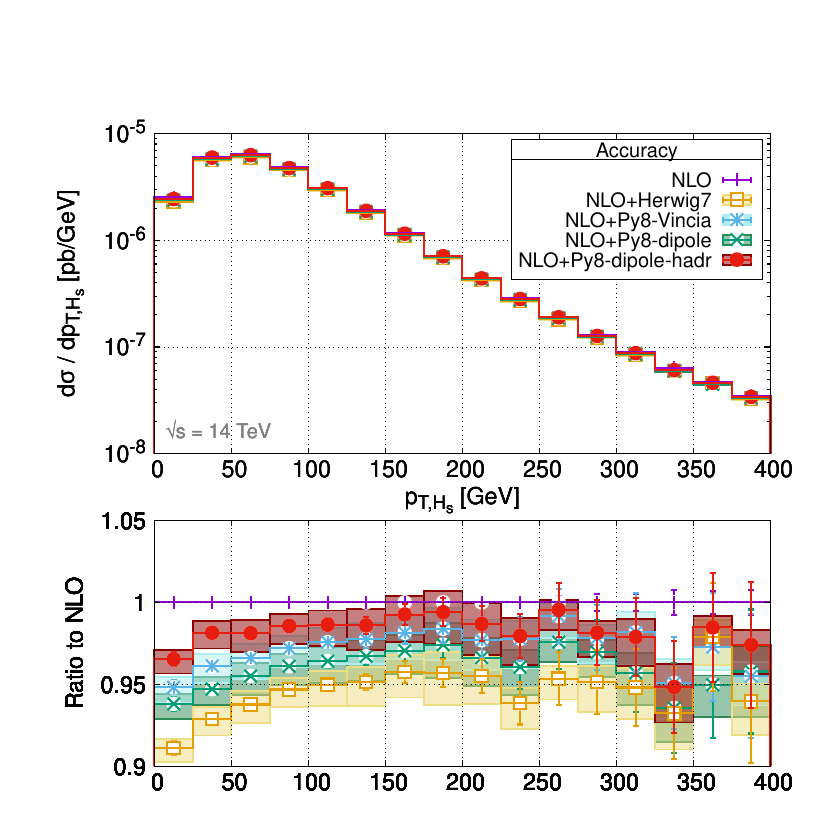}
\includegraphics[width=0.49\textwidth,page=2]{VBF-HH-PS-Plots_s.pdf} 
\caption{\label{fig:Hsoft-nlops} 
Transverse-momentum (left) and rapidity distributions (right) of the softer Higgs boson for the \vbfhh{} process as described in the text within the {\bf VBF cut} setup at NLO (purple) and at NLO+PS accuracy using \HERWIGS{} (orange) or  \PYTHIAE{} with the dipole shower without (green) and with (red) hadronisation/underlying event effects, and the \VINCIA{}  shower (blue). Their ratios to the respective NLO results are shown in the lower panels. Error bars indicate statistical uncertainties, uncertainty bands correspond to a 7-point variation around the central scale $\mu_0$ of each curve. 
} 
\end{figure}

\begin{figure}[pt!]
\centering
\includegraphics[width=0.49\textwidth,page=4]{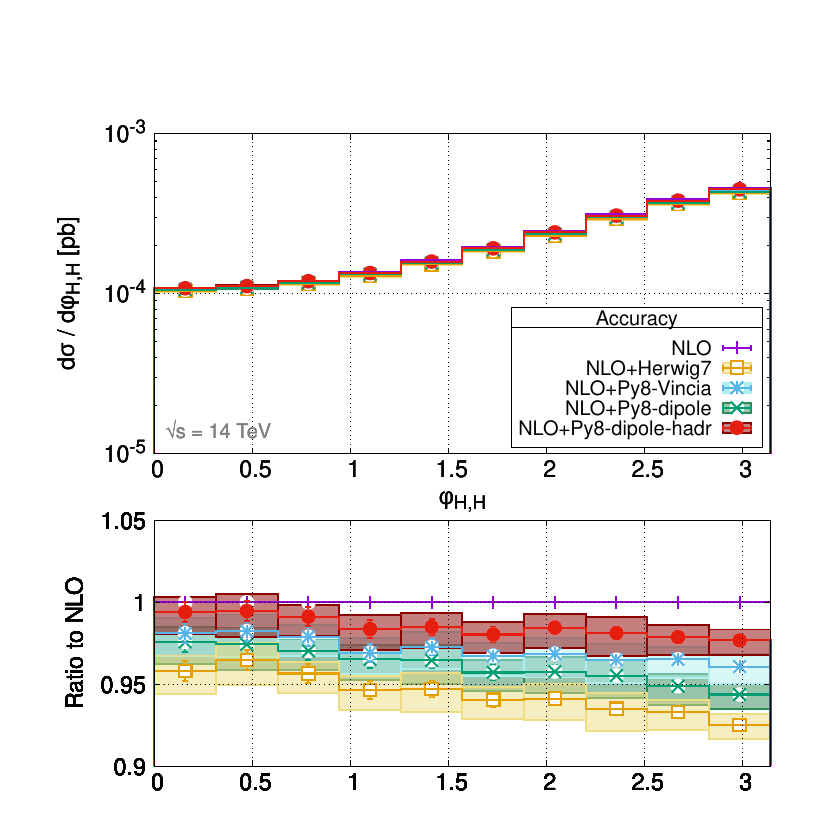}
\includegraphics[width=0.49\textwidth,page=3]{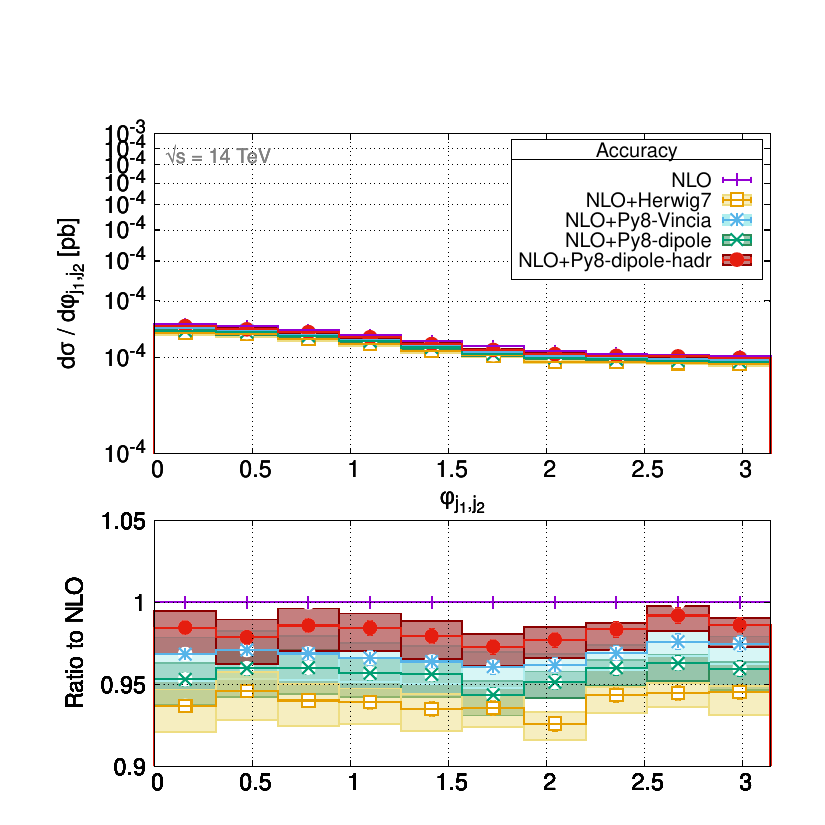} 
\caption{\label{fig:phi-nlops} 
Azimuthal angle separation distributions of the Higgs bosons (left) and the two taggings jets (right) for the \vbfhh{} process as described in the text within the {\bf VBF cut} setup at NLO (purple) and at NLO+PS accuracy using \HERWIGS{} (orange) or  \PYTHIAE{} with the dipole shower without (green) and with (red) hadronisation/underlying event effects, and the \VINCIA{} shower (blue). Their ratios to the respective NLO results are shown in the lower panels. Error bars indicate statistical uncertainties, uncertainty bands correspond to a 7-point variation around the central scale $\mu_0$ of each curve. 
} 
\end{figure}

Distributions of the tagging jets and the Higgs bosons do not exhibit a strong sensitivity to PS effects, as illustrated in Fig.~\ref{fig:Hsoft-nlops} for the transverse momentum and rapidity of the softer Higgs boson, and in Fig.~\ref{fig:phi-nlops} for the azimuthal angle separation of the two Higgs bosons and the two tagging jets. 
Differences between predictions obtained with the various PS programs are small, with the \VINCIA{} NLO+PS curves being closest to the fixed-order NLO results. Including hadronisation effects results in a slight increase of the NLO+PS predictions. 

\begin{figure}[pt!]
\centering
\includegraphics[width=0.49\textwidth,page=5]{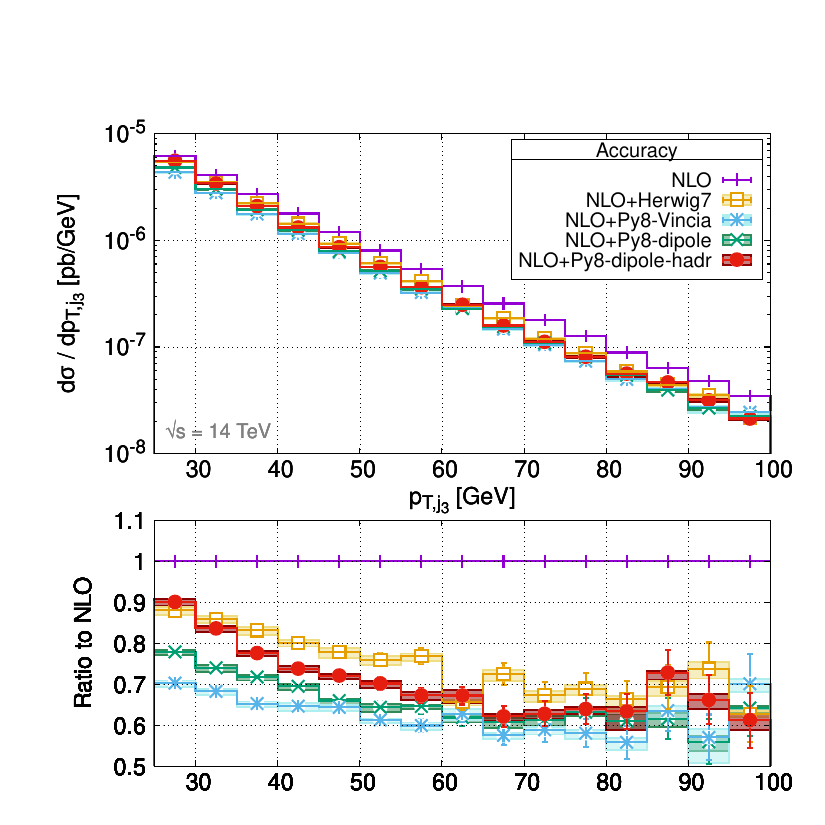}
\includegraphics[width=0.49\textwidth,page=6]{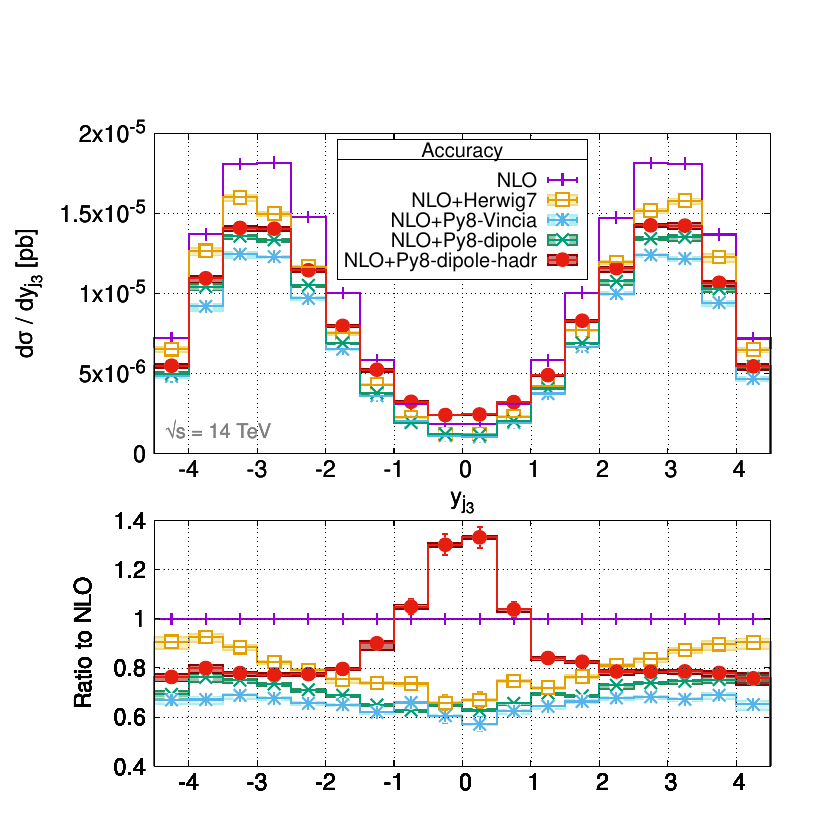} 
\caption{\label{fig:jet3} 
Transverse momentum (left) and rapidity distributions (right) of the third jet for the \vbfhh{} process as described in the text within the {\bf VBF cut} setup. The third jet is required to fulfill the cuts of Eq.~\eqref{eq:jets-sub}. Results are shown at NLO (purple) and at NLO+PS accuracy using \HERWIGS{} (orange) or \PYTHIAE{} with the dipole shower without (green) and with (red) hadronisation/underlying event effects, and the \VINCIA{} shower (blue). Their ratios to the respective NLO results are shown in the lower panels. Error bars indicate statistical uncertainties, uncertainty bands correspond to a 7-point variation around the central scale $\mu_0$ of each curve.
} 
\end{figure}

Larger PS effects are observed for distributions of non-tagging jets. In particular, Fig.~\ref{fig:jet3} 
illustrates the transverse-momentum and rapidity distributions of the third jet. Because of the cuts of Eq.~\eqref{eq:jets-sub}, only jets with transverse momenta larger than 25~GeV contribute to these observables. NLO+PS results are several tens of percent below the respective NLO predictions, and the shape of both distributions changes considerably when PS radiation is taken into account, in particular in the central-rapidity region. We note that these large differences between NLO+PS and fixed-order results are reduced by the inclusion of NNLO corrections in the fixed-order calculation, see Ref.~\cite{Jager:2025isz}. Even stronger changes in the shape of the third jet's rapidity distribution are obtained, if hadronisation effects are taken into account. In this case, the central-rapidity region tends to be significantly more populated.

\section{Conclusions}\label{sec:conclusion}
In this paper we have presented a systematic comparison of two tools for the simulation of the \vbfhh{} process at the LHC and HL-LHC. The two tools differ mainly in what contributions they include -- in particular \GOWIZ{} includes LO and NLO-QCD matrix elements for the full EW~$HHjj$ final state without resorting to any approximations. The \PBOX{} on the other hand resorts to the simpler VBF approximation. We found that for scenarios that are typically used in VBF analyses, and which are designed to suppress contributions beyond the VBF approximation, these tools provide similar predictions at LO and NLO-QCD accuracy both within the SM and in the presence of anomalous Higgs couplings. However, for more inclusive cuts the two predictions start to differ as expected.

This highlights the complementarity of the \GOWIZ{} and \PBOX{} for more general applications:  \GOWIZ{} includes the full LO and NLO-QCD matrix elements which is of relevance in inclusive selection scenarios. In principle, NLO results matched to a parton shower can also be obtained, but we leave this to future work. The \PBOX{} implementation instead uses the VBF approximation but includes an interface to PS generators such as \PYTHIA{} or \HERWIG{} using the \POWHEG{} formalism and thus directly provides NLO+PS results. 

\section*{Acknowledgements}

This work was done on behalf of the LHC Higgs Working Group. The research of PB, GH and MH is supported by the Deutsche Forschungsgemeinschaft (DFG, German Research Foundation) under grant 396021762 - TRR 257. JB is supported in parts by the Federal Ministry of Technology and Space (BMFTR) under grant number 05H24VKB. The authors also acknowledge support by the state of Baden-Württemberg through bwHPC and the DFG through grant no INST 39/963-1 FUGG. 

\bibliography{vbf_hh}

\end{document}